\documentclass[preprintnumbers,amsmath,amssymb,twocolumn,prl,showpacs]{revtex4-1}

\usepackage{graphicx,comment}

\usepackage{bm}
\usepackage{hyperref}
\usepackage{mathtools}

\usepackage{algorithm}
\usepackage{setspace}

\def\bb{\vv{b}'}

\def\bb{\bar{b}}

\def\cO{\mathcal{O}}

\def\Tr{\mathrm{Tr}}

\def\one{{\mathchoice {\rm 1\mskip-4mu l} {\rm 1\mskip-4mu l} {\rm
1\mskip-4.5mu l} {\rm 1\mskip-5mu l}}}
\newcommand{\ket}[1]{\left| #1\right\rangle}        
\newcommand{\sket}[1]{| #1\rangle}  
\newcommand{\bra}[1]{\left\langle #1\right|}        
\newcommand{\sbra}[1]{\langle #1|}
\newcommand{\ketbra}[1]{| #1 \rangle \! \mspace{2mu}\langle #1| }

\usepackage[dvipsnames]{xcolor}

\begin{document}

\title{Quantum algorithms for systems of linear equations \\inspired by adiabatic quantum computing}

\author{
Yi\u{g}it Suba\c{s}\i}
\affiliation{Theoretical Division, Los Alamos National Laboratory, Los Alamos, NM 87545, US.}

\author{
Rolando D. Somma}
\affiliation{Theoretical Division, Los Alamos National Laboratory, Los Alamos, NM 87545, US.}

 \author{Davide Orsucci}
\affiliation{Department of Theoretical Physics, University of Innsbruck, Innsbruck, Austria.}

\date{\today}

\begin{abstract}
We present two  quantum algorithms based on evolution randomization, a simple variant of adiabatic quantum computing, to prepare a quantum state $\ket x$ that is proportional to the solution of the system of linear equations $A \vec{x}=\vec{b}$.
The time complexities of our algorithms are $O(\kappa^2 \log(\kappa)/\epsilon)$ and $O(\kappa \log(\kappa)/\epsilon)$, where $\kappa$ is the condition number of $A$ and $\epsilon$ is the precision. 
Both algorithms are constructed using families of Hamiltonians that are linear combinations of products of $A$, the projector onto the initial state $\ket b$, and single-qubit Pauli operators.
The algorithms are conceptually simple  and easy to implement. They are not obtained from equivalences between the gate model and adiabatic quantum computing.
They do not use phase estimation or variable-time amplitude amplification,
and do not require large ancillary systems.
We discuss a gate-based implementation via Hamiltonian simulation and prove that our second algorithm is almost optimal in terms of $\kappa$.
Like previous methods, our techniques yield an exponential quantum speedup under some assumptions.
Our results emphasize the role of Hamiltonian-based models of quantum computing for the discovery of important algorithms.
\end{abstract}
           
 
\maketitle

\paragraph{\textbf{Introduction.}}
\label{sec:intro}
Recently, there has been significant interest in quantum algorithms to solve various linear algebra problems ~\cite{HHL09,Pra08,Amb12,CKS17,CGJ18}, as quantum computers can implement certain linear transformations more efficiently
than their classical counterparts. Such algorithms may find applications in a wide range of topics, including machine learning~\cite{WBL12,LMR13,LMR14}, graph problems~\cite{Wan13}, solving differential equations~\cite{Ber14}, and physics problems~\cite{CJS13,CS16}. 
A main example is the algorithm of Harrow, Hassidim, and Lloyd (HHL) of Ref.~\cite{HHL09} for the so-called quantum linear systems problem (QLSP), where the goal is to prepare a quantum state $\ket x$ that is proportional to the solution of a system of linear equations $A \vec x= \vec b$. 
If the $N \times N$ matrix $A$ and $N$-dimensional vector $\vec b$ are sparse, and for constant precision, the complexity of the algorithm in Ref.~\cite{HHL09} is polynomial in $\log N$ and $\kappa$, where $\kappa$ is the condition number of $A$.
In contrast, classical algorithms to invert matrices are of complexity polynomial in $N$, suggesting that quantum computers would be able to solve certain problems related to  systems of linear equations exponentially faster than classical computers. Improvements of the HHL algorithm can be found in Refs.~\cite{Amb12,CKS17,CGJ18}.

The referenced algorithms are described in the standard gate-based model of quantum computing, where quantum states are prepared by applying a sequence of elementary (e.g., two-qubit) gates to some initial state. 
However, Hamiltonian-based alternatives to the gate-based model exist, such as adiabatic quantum computing (AQC)~\cite{FGGS00}. 
One advantage of considering these other alternatives is that new and simple quantum algorithms can be found, even if such algorithms will ultimately be implemented using a different but equivalent model.

In AQC, for example, the computation is performed by smoothly changing the parameters of a Hamiltonian that evolves a quantum system. The adiabatic theorem asserts that if the continuously related eigenstates remain non-degenerate and the Hamiltonians change sufficiently slowly, then the evolved state is sufficiently close to the  eigenstate of the final Hamiltonian~\cite{Mes99}. Such an eigenstate encodes information about the solution to a problem; in our case the final eigenstate would be $\ket x$ (or $\ket \phi \otimes \ket{x}$ if ancillas are used).
A  closely related method is the randomization method (RM) described in Ref.~\cite{BKS09}. 
Both, AQC and RM are examples of eigenpath traversal~\cite{BKS10}.
Nevertheless, an advantage of the RM with respect to AQC is that better convergence guarantees can sometimes be obtained, as shown in Refs.~\cite{SBB08,CXS14}.

In this paper, we develop two simple quantum algorithms that solve the QLSP based on the RM.
To this end, we construct families of simple Hamiltonians whose continuously related eigenstates connect $\ket b$, the quantum state proportional to $\vec b$, with $\ket x$. 
The average evolution times of our algorithms, i.e.\ the time complexities, are nearly order $\kappa^2$ and $\kappa$, respectively. Here $\kappa$ is the condition number of $A$. 
Additionally, the time complexities of both algorithms are linear in $1/\epsilon$, where $\epsilon$ is a precision parameter. 
In contrast to previous approaches,
our algorithms do not use any form of phase estimation, amplitude amplification, or function approximation, thus reducing the number of ancillary qubits significantly.

Our first quantum algorithm solves the QLSP
by preparing the lowest-energy states of a family of Hamiltonians,
whereas our second algorithm achieves this by
preparing energy states that lie exactly at the middle of the spectrum, i.e., excited states.
Our second algorithm is noteworthy in that it is almost optimal, having time complexity almost linear in $\kappa$.

Although the Hamiltonians are simple, actual implementations of our algorithms on analog quantum computing devices may be impractical.
However, the quantum algorithms could still be implemented in the gate-based model by
using the Hamiltonian simulation results of Refs.~\cite{BAC07,BCC+15,LC17}.
This will require oracle access to the matrix $A$ as well as a procedure to prepare the state $\ket b$.
A resulting gate-model algorithm for the QLSP following this procedure will be  nearly optimal according to Ref.~\cite{HHL09}.
That is, like Refs.~\cite{Amb12,CKS17},
the query complexity is almost linear in $\kappa$, a
quadratic improvement over that of the HHL algorithm.
We give more specifics below.

\paragraph{\textbf{Quantum linear systems problem.}}
\label{sec:QLSP}
The QLSP  in Refs.~\cite{HHL09,Amb12,CKS17} is stated as follows. We are given an $N \times N$ Hermitian matrix $A$ and a vector $\vec b=(b_1,\ldots,b_N)^T$, with $N=2^n$. The goal is to prepare an $\epsilon$-approximation of a quantum state
\begin{align}
\label{eq:exactstate}
\ket x := \frac{\sum_{j=1}^N x_j \ket j }{  \sqrt{ \sum_{j=1}^N |x_j|^2 }} = \frac{(1/A) \ket b} {\| (1/A)\ket b\|} \;,
\end{align}
where $\vec x = (x_1,\ldots, x_N)^T$ is the solution to the linear system $A \vec x = \vec b$, $\ket b \propto \sum_{j=1}^N b_j \ket j$, and $0<\epsilon < 1$ is a precision parameter. We assume that $A$ is invertible, having condition number $\kappa < \infty$, and $\|A\| \le 1$. The approximated state $\ket{\tilde x}$ satisfies $\| \ket{\tilde x} - \ket x \| \le \epsilon$.
Here, we consider a slightly modified version of this problem
where the goal is to prepare a mixed state $\rho_x$ such that  the trace distance satisfies
\begin{align}
\label{eq:TD}
 \frac 1 2 \Tr   \left | \rho_x - \ketbra x \right| \le \epsilon \;.
\end{align}
Note that this modified version is adequate since the ultimate
purpose of the QLSP is for obtaining expectation values of observables. Thus, both $\ket{\tilde x}$ and $\rho_x$ will provide same-order approximations for such calculations.

\paragraph{\textbf{Algorithm evolving on ground states.}}
\label{sec:GSAlg}
We first define the family of Hamiltonians
\begin{align}
\label{eq:mainH}
	H(s) \; & := \;
	A(s) \, P_{\bb}^\perp \, A(s) \; .
\end{align}
Here, $A(s) := (1-s)Z \otimes \one + s X \otimes A$, $\sket{\bb} := \ket{+,b}$, $P_{\bb}^\perp := \one - \ketbra{\bb}$, and $s \in [0,1]$ is a parameter. 
$X$ and $Z$ are single-qubit Pauli operators.
These Hamiltonians act on a Hilbert space of dimension $2N$, i.e., the space of $A$ plus one ancilla qubit. 
The reason for using an ancilla is to
guarantee that $A(s)$ is
invertible for all $s$. We introduce the family of states
\begin{align}
    \label{eq:x(s)}
	\ket{x(s)} \; := \; \frac{1/(A(s)) \ket{\bb}} {\| 1/(A(s)) \ket{\bb} \|}\;,
\end{align}
which satisfy $H(s)\ket{x(s)} = 0$. In Supp.\ Mat.\ we show that $\ket{x(s)}$ is the unique ground state of $H(s)$
and the energy gap satisfies $\Delta(s) \ge \Delta^*(s):=(1-s)^2 + (s/\kappa)^2$. As $s$ is increased from $0$ to $1$, the ground state continuously changes from $\ket{x(0)}=\ket{-,b}$ to $\ket{x(1)}=\ket{+,x}$. Exact preparation of $\ket{x(1)}$ implies exact preparation of the target state $\ket{x}$ by discarding the ancillary qubit. 

In the case $A > 0$, we can opt for the simpler choice $A(s) := (1-s)\one + s A$, and still have $A(s)$ non-singular for all $s$. Then,  $\ket{x(s)} \propto A(s)^{-1} \ket{b}$ is the unique ground state of $H(s)$. 
The following analysis is for general $A$.

\paragraph{\textbf{Randomization method.}}
\label{sec:RM}
The details
of the RM as well as its complexity analysis can be found in Ref.~\cite{BKS09}.
Here, we mainly study and describe how to use the RM to solve the QLSP. 
Roughly, the method can be viewed as a version of AQC, where
the parameter $s$ is changed discretely rather than 
 continuously,
and the Hamiltonian evolution is for a random time.
This process effectively simulates an approximate projective measurement of the desired ground state (or any other eigenstate).
It then allows to make transformations
within the ground states (eigenstates) of the Hamiltonians.
The time complexity of the RM in general is
$O(L^2/(\epsilon \Delta))$, where $L$ is the so-called path length (which we define later), and $\Delta$ is the minimum gap of the Hamiltonians.
We observe that the dependence on $\Delta$ is optimal~\cite{CXS14}, while general bounds for AQC provide a worse time complexity of $O(1/\Delta^3)$~\cite{JRS07}.
This observation is key to achieve our results.
Then, obtaining the actual time complexity for the QLSP requires
studying the properties of the Hamiltonians $H(s)$ and eigenstates $\sket{x(s)}$.
With this information, we can find discrete values
of $s$ as well as values for the evolution times needed to implement the RM.

The full complexity analysis for the QLSP is given in Supp.\ Mat.. According to Refs.~\cite{BKS09,BKS10,CXS14}, to obtain the discrete values of $s$,
it is convenient to work with a ``natural'' parametrization $s(v)$.
This is defined so that the norm of the rate of change of the eigenstate
with respect to $v$ 
can be bounded by a constant. 
We find that a natural parametrization for this case is
\begin{align}
	s(v) \; := \;
	\frac{
	e^{v\frac{ \sqrt{1+\kappa^2}}{\sqrt 2 \kappa} } + 
	2 \kappa^2 - 
	\kappa^2 e^{-v\frac{ \sqrt{1+\kappa^2}}{\sqrt 2 \kappa} }}
	 {2(1+\kappa^2)} \;.
\end{align}
Here, $v_a \le v \le v_b$, with
\begin{align}
    \label{eq:va}
	v_a & \; := \; 
	\frac{\sqrt 2 \kappa}{ \sqrt{1+\kappa^2}} 
	\log(\kappa \sqrt{1+\kappa^2} - \kappa^2) \; , \\
	\label{eq:vb}
	v_b & \; := \;
	\frac{\sqrt 2 \kappa}{ \sqrt{1+\kappa^2}} 
	\log( \sqrt{1+\kappa^2} +1) \;.
\end{align}

The discrete values $s^j=s(v^j)$ are obtained from discrete values of $v$,
 which are evenly distributed in $q$ points as $v_a < v^1 < v^2 < \ldots < v^q=v_b$.
 Here, $v^j=v_a+j \delta$ ($j=1,\ldots, q$) and 
 $s^0=s(v_a)=0$, $s^q=s(v_b)=1$. The number of steps of the RM is $q=\Theta(\log^2(\kappa)/\epsilon)$, and $\delta=(v_b-v_a)/q$. 
 The choice of $q$ implies
  \begin{align}
 \label{eq:meas}
1-| \langle x(s^j) \sket{x(s^{j+1})}|^2 = O(\epsilon/q) \;.
 \end{align}
That is, a sequence of $q$ 
projective measurements of
 $\sket{x(s^j)}$, starting from $\sket{x(0)}$, will produce $\sket{x(1)}$ with probability $1-O(\epsilon)$. These measurements are simulated
 by evolution randomization.
 
Our algorithm is as follows.
At each step $j=1,\ldots,q$,
we evolve with the Hamiltonian $H(s^j)$ for a random time $t^j$. The evolution time can be sampled from the uniform distribution $t^j\in[0,2\pi/\Delta^*(s^j)]$~\cite{BKS09,CXS14} and satisfies
 $\langle  t^j  \rangle = \pi/(\Delta^*(s^j))$.
The time complexity of this algorithm is  $T := \sum_{j=1}^q \langle  t^j  \rangle$
and in Supp. Mat. we show
\begin{align}
T =O \left( \kappa^2 \log(\kappa)/\epsilon \right).
\end{align}
Note that, in each run, the overall evolution time
is always bounded by $2T$.

Our first algorithm then uses the RM
to prepare a mixed state $\rho_x$
that satisfies Eq.~\eqref{eq:TD}, after discarding the ancilla.
The time complexity is almost quadratic in $\kappa$.
The pseudocode for the algorithm is shown below.
\begin{algorithm}[H]
\renewcommand{\thealgorithm}{}
\floatname{algorithm}{\hspace{3.3cm} Algorithm}
\setstretch{1.25}
\vspace{1mm}
Given condition number $\kappa$ and precision $\epsilon$:\\
-- Compute $v_a$ and $v_b$. Set
$q\! =\! \Theta(\log^2(\kappa) /\epsilon)$,  $\delta\! =\! (v_b\!-\!v_a)/q$ \\
-- For $j=1,\dots,q$, let $v^j\! =\! v_a\! +\! j \delta$, $s^j=s(v^j)$,
and $t^j$ be sampled from the uniform distribution $\left[ 0,2\pi/\Delta^*(s^j) \right]$ \\
-- Apply $e^{-it^qH(s^q)} \ldots e^{-it^1H(s^1)}\!$ to  $\sket{\bar{b}}$, discard the ancilla
\\
\vspace{-3mm}
 \caption{}
\end{algorithm}

\paragraph{\textbf{Spectral gap amplification.}}
One way to improve the time complexity of the first algorithm is by considering other families of Hamiltonians where the relevant spectral gap is larger than
that of $H(s)$. This idea was considered in Ref.~\cite{SB13} and resulted in various polynomial quantum speedups for quantum state preparation. 
A quadratic spectral gap amplification is indeed possible when the Hamiltonians satisfy a so-called frustration free property. Very roughly, a possible Hamiltonian with an amplified gap can be interpreted as the square root of the frustration-free Hamiltonian. A zero eigenvalue remains zero and an eigenvalue $\lambda>0$ is transformed into eigenvalues $\pm \sqrt \lambda$. ($\sqrt \lambda \gg \lambda$ if $\lambda\ll 1$.) To avoid additional complexity overheads, the Hamiltonians with an amplified gap must satisfy certain constrains related to the difficulty of their simulation. We refer to~\cite{SB13} for details.

Motivated by these results, we now consider another family 
of Hamiltonians for solving the QLSP using the RM.
This family is given by
\begin{align}
    \label{eq:sqrtH}
    H'(s):=\sigma^+ \otimes A(s)P_{\bb}^\perp  +\sigma^- \otimes P_{\bb}^\perp A(s) \;,
\end{align}
where $\sigma ^{\pm} = (X \pm i Y)/2$ are single-qubit (raising and lowering) operators, 
and $s \in [0,1]$. We note that $H'(s)$ acts on a Hilbert space of 
dimension $4N$. Then
\begin{align}
\label{eq:SGA2}
    \left( H'(s) \right)^2 = \begin{pmatrix} H(s) & 0 \cr 0 & 
    P_{\bb}^\perp (A(s))^2  P_{\bb}^\perp \end{pmatrix}\;,
\end{align}
where each block of the matrix is of dimension $2N \times 2N$. Using $B(s) := A(s)P_{\bb}^\perp$, the blocks on the diagonal of Eq.~\eqref{eq:SGA2} can be written as $B(s)^\dag B(s)$ and $B(s)B(s)^\dag$, 
and thus have the same spectrum. Consequently, the eigenvalues of $H'(s)$ are $\{0,0,\pm \sqrt{\gamma_1(s)}, \ldots, \pm \sqrt{\gamma_{2N-1}(s)} \}$, where $\gamma_j(s) > 0$ are the nonzero eigenvalues of $H(s)$. The subspace of $H'(s)$ of eigenvalue zero is spanned by $\{\ket 0 \otimes \sket{x(s)},\ket 1 \otimes \sket \bb \}$.

In contrast to the first algorithm
that aimed at preparing the ground state
of $H(s)$,
we  now aim at preparing one
of the two eigenstates of zero eigenvalue of $H'(s)$
that lies exactly at the middle of the spectrum.
Nevertheless, the RM can be used to prepare
any eigenstate as long as it is separated by a nonzero
spectral gap from the other eigenstates.
One may wonder if the double degeneracy is a problem
for this case. The answer is negative as the 
Hamiltonian does not allow for transitions
between the two eigenstates, that is,
$\bra 0 \otimes \sbra {x(s)} H'(s') \ket 1 \otimes \sket{\bar b}=0$.
If we initialize our quantum computer
in $\ket 0 \otimes \sket{x(0)}$, a sequence of perfect projective
measurements of the eigenstates of $H'(s)$ at
sufficiently close points will allow us to prepare
 $\ket 0 \otimes \sket{x(1)}$
with sufficiently high probability.
The relevant spectral gap is now bounded by $\sqrt{\Delta^*(s)}>0$.

The eigenstate 
$\ket 0 \otimes \sket{x(s)}$
has similar properties as $\ket{x(s)}$:
the path length and norm of the rate of change
are the same. Then, our second algorithm can be constructed 
by using the same discretization points $s^j$
that were used for the first algorithm.
At each step, we now need to evolve
with the Hamiltonian $H'(s^j)$ for a random time
$t^j$. This time can be sampled from the uniform distribution
$t^j \in [0, 2 \pi/\sqrt{\Delta^*(s^j)}]$. The time
complexity of this algorithm is $T:=\sum_{j=1}^q \langle t^j \rangle$ and,
in Supp. Mat., we show
\begin{align}
\label{eq:alg2cost}
 T=O \left(  \kappa \log(\kappa)/ \epsilon \right) \;.
\end{align}

After discarding the two ancilla qubits,
the final state is $\rho_x$ and satisfies Eq.~\eqref{eq:TD}.
The time complexity of our second algorithm is then  
almost linear in $\kappa$. 
The pseudocode for this algorithm follows from the previous one by replacing $\Delta^*(s^j)$ with $\sqrt{\Delta^*(s^j)}$, $H(s)$ with $H'(s)$,  and $\sket{\bar{b}}$ with $\ket{0}\otimes\sket{\bar{b}}$, in the second and third lines.

\paragraph{\textbf {Simulation results.}}
\label{sec:sim}
We tested the validity of our quantum algorithms by performing numerical simulations. For this purpose, we randomly generated Hermitian matrices $A$ of dimension $N = 16$ that are $4$-sparse and $N=32$ that are $5$-sparse, both satisfying $\|A\|=1$. The generated matrices result in a range of values for the condition number $\kappa$. We post-selected matrices for which $\kappa \approx 10$ and $\kappa \approx 50$ (to within absolute error $10^{-3}$), for $N=16$ and $N=32$ respectively.
Similarly, we randomly generated 4-sparse and 5-sparse vectors for $\vec{b}$.  The parameters $s^j$ and $t^j$ were chosen according to the previous discussion and depend on $\kappa$ and $\epsilon$ (or $q$). In each execution, we prepare a pure quantum 
state that is not guaranteed to be $\epsilon$-close
to the pure eigenstate of the final Hamiltonian.
However, the expected error of the prepared pure states
from many repeated executions of the algorithms is indeed bounded by $\epsilon$.

We ran simulations for which the number of repetitions of our algorithms
were $n_{rep}=50$ and $n_{rep}=200$, respectively. 
For each case, we first construct a finite-sampling density matrix $(1/n_{rep})\sum_{i=1}^{n_{rep}}\ketbra{\psi_i}$. Here,  $\ket{\psi_i}$ is the pure state
output at the $i$'th repetition. Tracing out the ancilla qubits, we get a density matrix $\tilde{\rho}_x$ that describes the state of the system only. 
Note that $\tilde{\rho}_x$ is, in general, slightly different from $\rho_x$ of Eq.~\eqref{eq:TD}. However,
$\tilde{\rho}_x \rightarrow {\rho}_x$   in the limit of $n_{rep}\rightarrow \infty$.
The error computed in our numerical simulations is then the trace distance between $\tilde{\rho}_x$ and $\ketbra{x}$.

\begin{figure}
    \centering
    \hbox{\hspace{-2.0em} \includegraphics[scale=0.65]{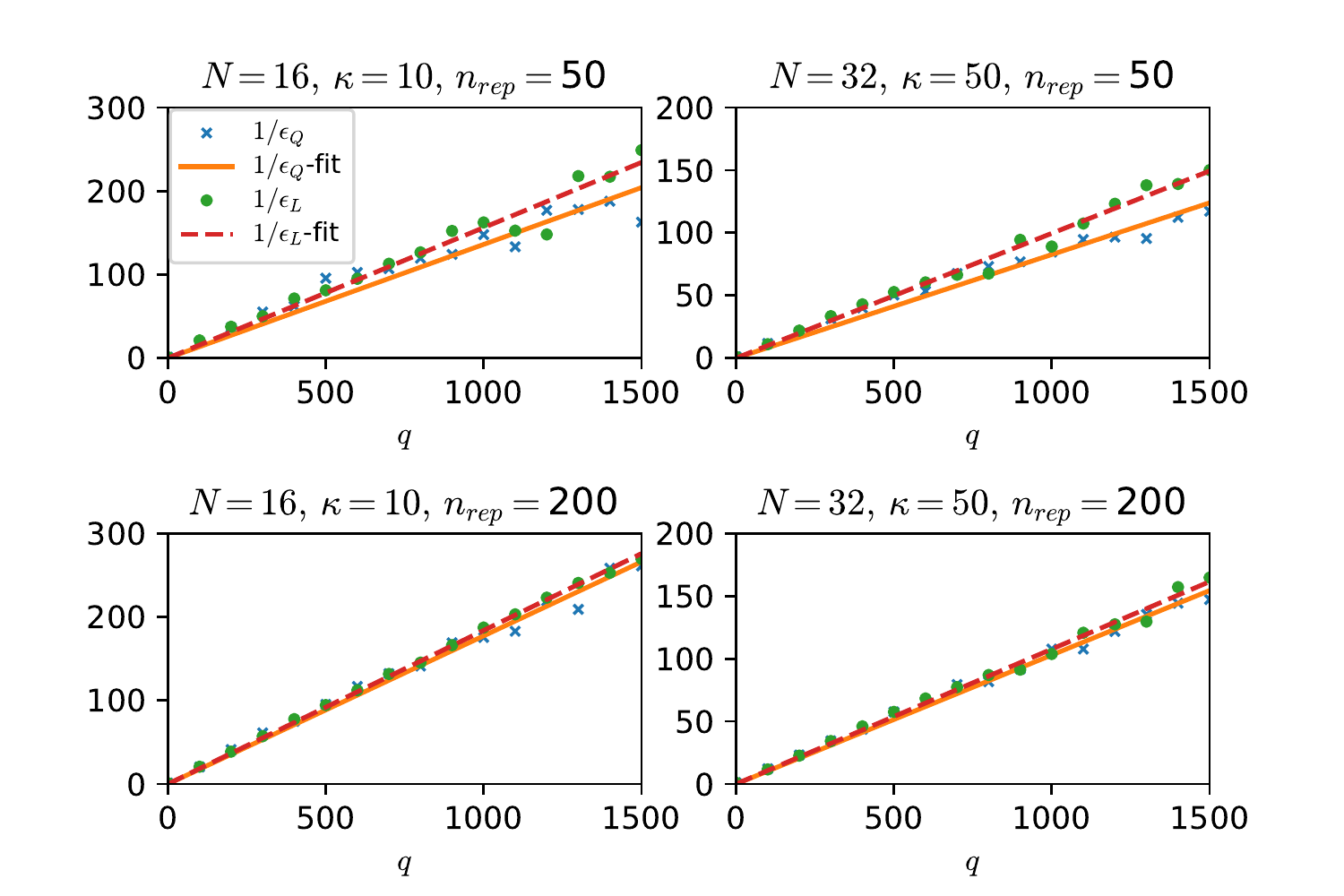} }
     \caption{The inverse of the error for the two quantum algorithms as a function of $q$, the number of steps in the RM. Subscript $Q$ refers to the quantum algorithm with complexity that is almost quadratic in $\kappa$ and $L$ to the quantum algorithm with complexity almost linear in $\kappa$.
    $n_{rep}$ is the number of repetitions of the of our algorithm.
    The results are for two randomly generated matrices $A$ with $N=16,\, \kappa \approx 10$, and $N=32,\, \kappa \approx 50$.
    }
    \label{fig:error_vs_q}
\end{figure}

In Fig.~\ref{fig:error_vs_q}, we show the dependence of the inverse of the error on the number of steps $q$.
While the results are for two particular matrices $A$ with
$\kappa \approx 10$ and $\kappa \approx 50$, other matrices
show similar results.
We observe that the inverse of the error for the two quantum algorithms,
denoted by  $\epsilon_Q$ and $\epsilon_L$ respectively, scales almost linearly with $q$.
The dispersion around the linear fit is
smaller for larger $n_{rep}$. 
The results are then in accordance with our theoretical analysis.

\paragraph{\textbf{Gate-based implementations.}}
\label{sec:gate-based}
Our algorithms are based on Hamiltonian evolutions
and can be implemented on a gate-based quantum computer
using a Hamiltonian simulation method. 
We focus on the method of Ref.~\cite{BCC+15},
which implements the truncated
Taylor series of the evolution operator.
It requires the Hamiltonian to be given as a linear combination
$\sum_l \alpha_l V_l$, where the $V_l$
are unitaries that are easy to apply and $\alpha_l>0$.
The $V_l$ are
applied $\tilde O(\tau)$
times, where
$\tau$ is the product of the evolution time
and $\sum_l \alpha_l$.
The $\tilde O$ notation hides logarithmic factors
in $\tau$.

Our second algorithm applies the evolution 
under $H'(s^j)$ for time $t^j$.
The main challenge is then to
find a decomposition of the Hamiltonian
in terms of unitaries.
For technical reasons, we consider another Hamiltonian
$\tilde H'(s^j)$, but whose evolution operator
mimics that of $H'(s^j)$. This Hamiltonian
is discussed in Supp. Mat..
It turns out that $\tilde H'(s)=\frac{d+1}{16} \sum_{l=1}^{32} V_l(s)$,
where $V_l(s)$ are unitaries.
As previous approaches for the QLSP~\cite{CKS17}, we assume access to a quantum oracle $\cO_A$ for the matrix $A$.
This oracle outputs the nonzero matrix elements and their indices,
for any row of $A$. We also assume access
to a (controlled) unitary $U_b$ that prepares the state $\ket b$
and the (controlled) $U^\dagger_b$, as in Refs.~\cite{HHL09,Amb12,CKS17}.
Each unitary $V_l(s)$ can be applied using, at most, a constant number
of $\cO_A$ and (controlled) $U_b$ and $U^\dagger_b$.
In addition, it may require $O(n)$ two-qubit gates --
see Supp. Mat..

In our construction, we have
$\tau=O(t^j d)$ if
the evolution time is $t^j$. 
Since our algorithm implements evolutions with $q$ Hamiltonians,
the total number of uses of  $\cO_A$ and (controlled) $U_b$ and $U^\dagger_b$,
or query complexity,
is then $\tilde O(Td)$, 
where $T$ is the total evolution time.
The number of additional two-qubit gates is
a multiplicative factor of order $n$
away from
the query complexity.

Substituting $T$ from Eq.~\eqref{eq:alg2cost}
gives the query complexity of our approach
as $\tilde O(\kappa d/\epsilon)$. 
In Ref.~\cite{HHL09}, it was shown that quantum
algorithms for the QLSP must have a query complexity
that is, at least, linear in $\kappa$. 
Then, the gate-based implementation following Ref.~\cite{BCC+15} is almost optimal.
Note that the query complexity of evolving
with $\tilde H'(s)$ is of the same order as that
of evolving directly with $A$, which is needed for the HHL algorithm.

\paragraph{\textbf {Discussion.}}
\label{sec:discussion}
We presented simple quantum algorithms  for solving the QLSP that were motivated by AQC and not the standard gate-based model. 
A nice feature about AQC and related models, such as the RM or general eigenpath traversal methods~\cite{BKS10}, is that the time complexity is typically dominated by a single quantity, i.e., the inverse of the minimum spectral gap of the Hamiltonians.
Then, the root of the quantum speedup is more clear in this case than in the gate-based model, allowing for algorithmic improvements by considering different Hamiltonians with larger spectral gaps. Another nice feature is that some problems are naturally reduced to preparing the eigenstate of a Hamiltonian, and eigenpath traversal methods are useful in that context. We showed that this is the case for the QLSP. 
Our results emphasize the importance of considering models of quantum computing, which go beyond the gate-based model, for discovering novel quantum algorithms -- see Ref.~\cite{SNK12} for another example.

The further significance of our results is as follows.
Previous algorithms for the QLSP \cite{HHL09,Amb12,CKS17} use three main subroutines: (i) Hamiltonian simulation, (ii) phase estimation or function approximation, 
and (iii) some form of amplitude amplification. 
The method of ``variable-time amplitude amplification'' is used in Refs.~\cite{Amb12,CKS17} to achieve near-optimal complexity in terms of $\kappa$. 
That method alone requires  $\Omega(\log(1/\epsilon)\log(\kappa/\epsilon)/\epsilon^2)$ and $\Omega(\log(\kappa)\log(\kappa/\epsilon))$ ancillary qubits, respectively, which become excessively large for large $\kappa$. 
In contrast, our algorithms use only Hamiltonian simulation (which has the same query complexity as in previous methods) thereby reducing the number of ancillary qubits significantly.
Our result additionally implies a significant reduction in the number of conditional operations to solve the QLSP, making our algorithms more attractive for implementations on quantum computers of smaller size.
To this point, our algorithm has already been used in Ref.~\cite{Wen18} to solve an 8-dimensional linear system on a 4-qubit NMR device, the largest 
dimension up to date.

The time complexity of our methods is linear
in $1/\epsilon$. 
This complexity can be improved to polylogarithmic
in $1/\epsilon$ using the fast methods for eigenpath traversal
of Ref.~\cite{BKS10}. These methods will provide a different way of obtaining
an exponential improvement in terms of precision with respect 
to the HHL algorithm, as in Ref.~\cite{CKS17}, 
but they  nevertheless require
phase estimation.

Last, it would be interesting to study if our results
can also impact classical methods for solving systems of linear  equations.

\paragraph{\textbf{Acknowledgements.}}
We thank Anirban Chowdhury for discussions.
Part of this material is based upon work supported by the U.S. Department of Energy, Office of Science, Office of Advanced Scientific Computing Research, Quantum Algorithms Teams program.
We also thank the Laboratory Directed Research and
Development Program at LANL for 
support.
DO acknowledges the support by the Austrian Science Fund (FWF) through the DK-ALM W1259-N27, the SFB FoQuS F4012, and the Templeton World Charity Foundation grant TWCF0078/AB46.

\bibliography{AdiabaticInverse}

\clearpage

\section{Supplementary Material}
\label{sec:suppmat}

We consider first the Hamiltonians $H(s)$ of Eq.~\eqref{eq:mainH}
and note that
\begin{align}
H(s) (1/A(s)) \sket \bb &= A(s) P_{\bar b}^\perp \sket \bb \\
\nonumber & =0 \;.
\end{align}
$A(s)$ is always invertible and its eigenvalues are $\pm \sqrt{(1-s)^2 + (s\lambda)^2}$,
where $\lambda$ is an eigenvalue of $A$.
We let $B(s)=A(s)P_{\bar b}^\perp$ and, since $H(s)=B(s)B^\dagger(s)$,
we obtain $H(s) \ge 0$. Also, the only
eigenstate of zero eigenvalue of $P_{\bar b}^\perp$ is $\sket \bb$,
implying that $\ket{x(s)}$ is the unique eigenstate of zero eigenvalue (ground state) of $H(s)$. Any other eigenvalue is positive
and bounded from above by 1, under the assumptions on $A$.

The Hamiltonian $H(s)$ has smallest eigenvalue zero
and second smallest eigenvalue $\Delta(s)$.
It can be written as $H(s)=A^2(s) -A(s) \ketbra \bb A(s)$.
Weyl's inequalities~\cite{Bha97} imply that the second smallest
eigenvalue of $H(s)$ can be lower bounded by the smallest
eigenvalue of $A^2(s)$ plus the second smallest eigenvalue of 
$-A(s) \ketbra \bb A(s)$, which is zero. 
It follows that the spectral gap of $H(s)$
satisfies
\begin{align}
\label{eq:gapp}
\Delta(s) \ge \Delta^*(s):=(1-s)^2+(s/\kappa)^2 \;, 
\end{align}
where $\Delta^*(s)$ is a bound on the square of the smallest eigenvalue of $A(s)$.

Under the assumption $\bra{x(s)}\partial_s x(s)\rangle=0$,
the rate of change $\| \sket{\partial_s x(s)} \|$ can be obtained as follows.
First,
\begin{align}
\label{eq:roc}
 \ket{ \partial_s x(s)} =\frac 1 {A(s)} (Z \otimes \one - X \otimes A) \ket{x(s)} +\beta(s) \ket{x(s)} \;,
 \end{align}
 where $\beta(s) \in \mathbb{R}$. Since Eq.~\eqref{eq:roc} is orthogonal to $\ket{x(s)}$, we obtain
 \begin{align}
  \ket{ \partial_s x(s)}= P_x^\perp \frac 1 {A(s)} (-Z \otimes \one + X \otimes A) \ket{x(s)} \;,
 \end{align}
 with $P_x^\perp = \one - \ketbra{x(s)}$ the orthogonal projector. Using simple norm properties,
 it follows that
 \begin{align}
 \label{eq:roc2}
\|  \ket{ \partial_s x(s)} \|& \le \sqrt 2 \| 1/A(s) \| \\
\nonumber
& \le \sqrt{\frac 2 {(1-s)^2 + (s/\kappa)^2}} \;.
 \end{align}
 The quantity $\| \sket{\partial_s x(s)} \|$ is useful to compute
 the path length of the eigenstate. If the assumption 
 $\bra{x(s)}\partial_s x(s)\rangle = 0$ is not satisfied,
 we need to redefine $\ket{x(s)}$ by introducing a phase factor $e^{i\phi(s)}$
 so that the assumption is satisfied. The calculation of $\ket{ \partial_s x(s)}$
 has now an extra term that is due to the rate of change of $\phi(s)$.
 Nevertheless, the bound of Eq.~\eqref{eq:roc2} still applies when this extra phase
 factor is introduced.
 
 \paragraph{\textbf{Natural parametrization.}} 
 A natural parametrization $v(s)$ is such that 
$ \|  \ket{ \partial_v x(s(v))} \|$ is upper bounded by a constant, e.g.\ 1.
This parametrization can be obtained from
Eq.~\eqref{eq:roc2} if we require
\begin{align}
\label{eq:appNatPar}
\partial_v s = \sqrt{\frac {(1-s)^2 + (s/\kappa)^2} 2} \;.
\end{align}
The solution is
\begin{align}
\label{eq:natpar}
s(v) = \frac{e^{v\frac{ \sqrt{1+\kappa^2}}{\sqrt 2 \kappa}  } + 2 \kappa^2 - \kappa^2 e^{-v\frac{ \sqrt{1+\kappa^2}}{\sqrt 2 \kappa}  }}  {2(1+\kappa^2)} \;.
\end{align}
This function increases monotonically in the region of interest.
We define
\begin{align}
v_a &= \frac{\sqrt 2 \kappa}{ \sqrt{1+\kappa^2}} \log(\kappa \sqrt{1+\kappa^2} - \kappa^2) \; , \\
v_b & = \frac{\sqrt 2 \kappa}{ \sqrt{1+\kappa^2}} \log( \sqrt{1+\kappa^2} +1) \; ,
\end{align}
and note that
\begin{align}
s(v_a)=0 \; ,\quad s(v_b)=1 \;,
\end{align}
so that the boundary constrains are satisfied.

In the following, we abuse the notation slightly
so that $f(v):=f(s(v))$. The notation will be clear from the context.

\paragraph{\textbf {Path length.}}
The path length
\begin{align}
L &:= \int_0^1 ds \; \| \ket{\partial_s x(s)} \| \\
\nonumber
& = \int_{v_a}^{v_b} dv \; \| \ket{\partial_v x(v)} \|
\end{align}
is a useful quantity for methods based on eigenpath traversal. 
From the natural parametrization of Eq.~\eqref{eq:appNatPar},
it follows that $\| \ket{\partial_v x(v)} \| \le 1$ and
\begin{align}
L & \le \int_{v_a}^{v_b} dv = v_b-v_a \;.
\end{align}

Additionally, we can use the fact that
$\kappa+1/(4\kappa)\le\sqrt{\kappa^2+1}\le \kappa+1$ to show
\begin{align}
\nonumber
    v_b-v_a &=\frac{\sqrt 2 \kappa}{\sqrt{1+\kappa^2}} \log
    \left( \frac{\sqrt{1+\kappa^2}+1}{\kappa \sqrt{1+\kappa^2}-\kappa^2} \right) \\
    \nonumber
    & \le \sqrt 2 \log \left( \frac{\kappa+2} {1/4}\right ) \\
    &\le \sqrt 2 \log(12 \kappa)=L^* \;.
\end{align}
Then, the upper bound $L^*$
on the path length depends logarithmically on the condition number.

\paragraph{\textbf{ Time complexity using $H(s)$.}}
The time complexity of our first algorithm is
\begin{align}
\label{eq:apptimecomp1}
T =\pi \sum_{j=1}^q 1/\Delta^*(v^j) \;.
\end{align}
According to Ref.~\cite{BKS09}, we can choose the number of steps $q$
to be $\Theta((L^*)^2/\epsilon)$.
We can multiply and divide Eq.~\eqref{eq:apptimecomp1}
by $\delta:=L^*/q$. As $\kappa$ gets larger, 
$\delta$ gets smaller, and we should be able to approximate $T$
by $(\pi/\delta)\int_{v_a}^{v_b} dv \; (1/\Delta^*(v))$.
More precisely, the function $1/\Delta^*(v)$ reaches its maximum value
of $\kappa^2+1$ at $v=v_M$, and $v_a \le v_M \le v_b$. We let $r$ be the integer in
$\{1,\ldots,q\}$ such that $v^r \le v_M \le v^{r+1}$. It is
then possible to show
\begin{align}
   & \delta \sum_{j=1}^r \frac 1 {\Delta^*(v^j)} \le  \int_{v_a}^{v^r-\delta} dv \; \frac 1 {\Delta^*(v+\delta)} + \delta \frac 1 {\Delta^*(v^r)} \; , \\
    & \delta \sum_{j=r+1}^q \frac 1 {\Delta^*(v^j)} \le \delta \frac 1 {\Delta^*(v^{r+1})} + 
     \int_{v^{r+1}}^{v_b} dv \; \frac 1 {\Delta^*(v)} \;.
\end{align}
Then,
\begin{align}
\label{eq:InvGapBound}
    \sum_{j=1}^q \frac 1 {\Delta^*(v^j)} \le \frac 1 \delta \int_{v_a}^{v_b} dv \; \frac 1 {\Delta^*(v)}
    + \frac 2 {\Delta^*(v_M)} \;.
\end{align}
The second term on the right hand side is $2(\kappa^2+1)$. To obtain the first
term on the right hand side, we evaluate the integral:
\begin{align}
\nonumber
    \int_{v_a}^{v_b} dv \; \frac 1 {\Delta^*(v)} &= \int_0^1 ds \; \frac{\sqrt 2}{[(1-s)^2+(s/\kappa)^2]^{3/2}} \\
    & = \sqrt 2 \kappa (1+\kappa) \;.
\end{align}
These bounds imply
\begin{align}
    T \le \pi \left[ \frac {\sqrt 2 \kappa (1+\kappa)} \delta + 2 (\kappa^2 +1) \right]
\end{align}
Finally, recalling that 
$\delta =\Theta(\epsilon/\log(\kappa))$ and $q=\Theta (\log^2(\kappa)/\epsilon)$, we obtain $T=O(\kappa^2 \log( \kappa)/\epsilon)$.

\paragraph{\textbf{Spectral gap amplification.}}
Consider the Hamiltonian $I(s):=P_{\bb}^\perp (A(s))^2 P_{\bb}^\perp$
that appears in one of the diagonal blocks of Eq.~\eqref{eq:SGA2}.
This can be written as $I(s)=B^\dagger(s)B(s)$. Then, if $\ket{\gamma(s)}$ is an eigenstate of $H(s)$
of eigenvalue $\gamma(s)>0$, $B^\dagger(s) \ket{\gamma(s)} \ne 0$ is an eigenstate
of $I(s)$ with the same eigenvalue. The eigenstate of eigenvalue zero of $I(s)$
is $\sket{\bb}$.

\begin{figure}[htb!]
    \includegraphics[width=0.38\textwidth]{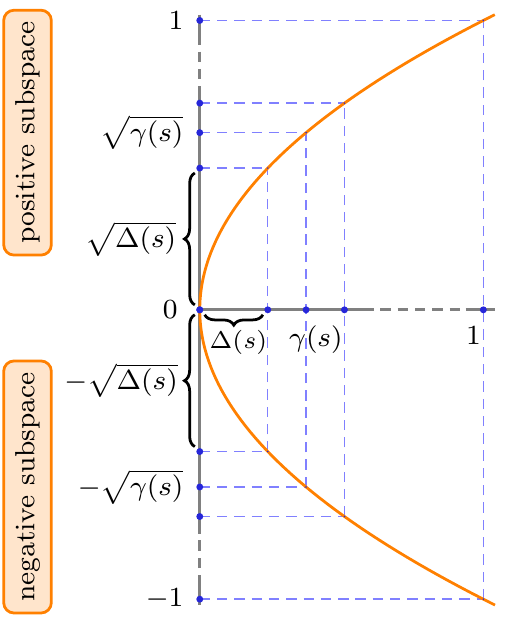}
    \caption{Schematic illustration of the spectral gap amplification method applied to the Hamiltonians $H(s)$. The positive (negative) subspace refers to the subspace
    of $H'(s)$ of positive (negative) eigenvalues. The horizontal axis
    shows the  eigenvalues of $H(s)$ and the vertical axis
    shows the eigenvalues of $H'(s)$.
    }
    \label{fig:SGA}
\end{figure}

Equation~\eqref{eq:SGA2} implies that the eigenvalues of $H'(s)$
are 0, $+\sqrt{\gamma(s)}$, and $-\sqrt{\gamma(s)}$. Moreover,
$(Z \otimes \one) H'(s) (Z \otimes \one)=-H'(s)$, so if $+\sqrt{\gamma(s)}$
is an eigenvalue of $H'(s)$ then $-\sqrt{\gamma(s)}$ is also an eigenvalue.
In summary, $H'(s)$ has a zero eigenvalue of degeneracy two and
the other eigenvalues are $\pm \sqrt{\gamma(s)}$ for each eigenvalue $\gamma(s)$
of $H(s)$.
This is illustrated in Fig.~\ref{fig:SGA}.

\paragraph{\textbf{Time complexity using $H'(s)$.}}
As described in the main text, the time complexity of 
our second algorithm is 
\begin{align}
    T = \pi \sum_{j=1}^q 1/\sqrt{\Delta^*(v^j)}.
\end{align}
Following Eq.~\eqref{eq:InvGapBound} for this case, we obtain
\begin{align}
    \sum_{j=1}^q \frac 1 {\sqrt{\Delta^*(v^j)}} \le \frac 1 \delta \int_{v_a}^{v_b} dv \; \frac 1 {\sqrt{\Delta^*(v)}} + \frac 2 {\sqrt{\Delta^*(v_M)}} \;.
\end{align}
The second term on the right hand side is $2 \sqrt{\kappa^2+1}$.
To obtain the first term on the right hand side, we evaluate the integral
\begin{align}
\nonumber
    \int_{v_a}^{v_b} dv \; \frac 1 {\sqrt{\Delta^*(v)}}&=
    \int_0^1 ds \; \frac{\sqrt 2}{(1-s)^2 + (s/\kappa)^2} \\ &
    =\pi \kappa/\sqrt 2 \;.
\end{align}
These bounds imply
\begin{align}
T \le \pi \left[ \frac{\pi \kappa}{\sqrt 2 \delta} + 2 \sqrt{\kappa^2+1} \right] \;,
\end{align}
and then $T=O(\kappa \log( \kappa)/\epsilon)$.

\paragraph{\textbf{Special Case: $A>0$.}}
In this case, we can define
\begin{align}
    A(s) &:= (1-s) \one + s A\, \label{eq:appA(s)}\; .
\end{align}
The important property is that $A(s)$ is invertible for $s \in [0,1]$, and we achieve this without using an ancilla qubit.

The Hamiltonians for our two algorithms can be modified by simply replacing $A(s)$ in Eqs.~(\ref{eq:mainH}) and~(\ref{eq:sqrtH}) by the one in Eq.~\eqref{eq:appA(s)}. The same bounds for the spectral gap and path length that apply for general $A$ also hold for this case. Then, we can use the same natural parametrization of Eq.~\eqref{eq:natpar}. As a result, the QLSP for $A>0$ can be solved with the same time complexity as in the general case, albeit using one fewer ancilla qubit and even a simpler Hamiltonian.

\paragraph{\textbf{Gate-based implementations.}}
We provide the details of the implementation of the algorithm based on the Hamiltonian of Eq.~\eqref{eq:sqrtH} 
using a gate-based quantum computer. 
Our goal is to use the Hamiltonian simulation method of Ref.~\cite{BCC+15},
which requires expressing the Hamiltonian as a linear combination of unitaries.
$A(s)$ can be written as a linear combination of unitaries using a version of Szegedy walks that applies to Hermitian matrices~\cite{Sze04,BCK15}. 
$A$ is of dimension $N$, $n=\log_2 N$, and $A(s)$ is of dimension $2N$.
We define
unitary operations $U_x$, $U_y$, and $S$ that act as follows:
\begin{widetext} 
\begin{align}
    U_x \ket{j} \ket{0} \ket{0} \ket{0} &= \frac{1}{\sqrt{d+1}}\sum_{i\in F_j} \ket{j} \ket{i} \ket{0} \left( \sqrt{A(s)_{ji}^*} \ket{0} + \sqrt{1-\vert A(s)_{ji} \vert} \ket{1}\right) \;, \\
    U_y \ket{0} \ket{j'} \ket{0} \ket{0} &= \frac{1}{\sqrt{d+1}}\sum_{i'\in F_{j'}} \ket{i'} \ket{j'}  \left( \sqrt{A(s)_{i'j'}} \ket{0} + \sqrt{1-\vert A(s)_{i'j'} \vert} \ket{1}\right)\ket{0} \;, \\
    S \ket{j}\ket{j'}\ket{\cdot}\ket{\cdot} &= \ket{j'}\ket{j}\ket{\cdot}\ket{\cdot} \; .
\end{align}
\end{widetext}
The first two registers have $n+1$ qubits each,
the last two registers have a single qubit each, and 
$F_j$ is the set of indices $i$ for which $A(s)_{ji}$ is nonzero.
It follows that
\begin{align}
\nonumber
    A(s) \otimes \ketbra{\tilde 0}&= (d+1) \; \ketbra{\tilde{0}}  U_x^\dagger U_y S \ketbra{\tilde{0}} \\
    \label{eq:LCUofA}
    &= \frac{d+1}{4}\left( \one - e^{i\pi P }\right) U_x^\dagger U_y S \left( \one - e^{i\pi P }\right) \; ,
\end{align}
where we defined $\sket{\tilde{0}} := \ket{0}\ket{0}\ket{0}$ for the state of the last three registers and $P= \ketbra{\tilde 0}$.

We assume access to an oracle for $A$ that acts as
\begin{align}
     \ket{j}\ket{i}\ket{z} & \rightarrow \ket{j}\ket{i}\ket{z\oplus A_{ji}} \;,\\
      \ket{j}\ket{l} & \rightarrow \ket{j}\ket{f(j,l)} \;.
\end{align}
Here, $j$ and $i$ label the row and column of $A$,
respectively, so that $j,i \in\{1,\ldots,N\}$.
 $f(j,l)$ is the column index of the $l$'th nonzero element of $A$ in row $j$. We refer to this oracle as $\mathcal{O}_A$. 
 From the oracle $\mathcal{O}_A$ it is straightforward to implement an oracle  $\mathcal{O}_{A(s)}$ for the $(d+1)-$sparse matrix $A(s)$. 
$\mathcal{O}_A$ is the same as that used in previous works for the QLSP
and Hamiltonian simulation; cf.
Refs.~\cite{CKS17,BCK15}.

$U_x$ can then be implemented in five steps as follows:
\begin{widetext}
\begin{align}
      \ket{j}\ket{0}\ket{0}\ket{0} \ket{0} &\xrightarrow[\text{Hadamards}]{\log(d+1)} \frac{1}{\sqrt{d+1}}\sum_{l=0}^{d} \ket{j}\ket{l}\ket{0}\ket{0} \ket{0}\\
    &\xrightarrow[]{\mathcal{O}_{A(s)}} \frac{1}{\sqrt{d+1}} \sum_{i\in F_j} \ket{j}\ket{i}\ket{0}\ket{0}\ket{0} \\
     &\xrightarrow[]{\mathcal{O}_{A(s)}} \frac{1}{\sqrt{d+1}} \sum_{i\in F_j} \ket{j}\ket{i}\ket{A(s)_{ji}}\ket{0}\ket{0} \\
     &\xrightarrow[]{~~M~~} \frac{1}{\sqrt{d+1}} \sum_{i\in F_j} \ket{j}\ket{i}\ket{A(s)_{ji}}\ket{0}\left( \sqrt{A(s)_{ji}^*}\ket{0} + \sqrt{1-\vert A(s)_{ji}\vert}\ket{1} \right) \\
    &\xrightarrow[]{\mathcal{O}_{A(s)}} \frac{1}{\sqrt{d+1}} \sum_{i\in F_j} \ket{j}\ket{i}\ket{0} \ket{0}\left( \sqrt{A(s)_{ji}^*}\ket{0} + \sqrt{1-\vert A(s)_{ji}\vert}\ket{1} \right) \;.
\end{align}
\end{widetext}
The third register is used to temporarily store the matrix elements of $A(s)$ and is discarded at the end.
$U_x$ can then be implemented with $O(1)$ queries to $\mathcal{O}_A$. 
The complexity of the fourth step, where the operation $M$ is implemented,
is defined to be $C_M$. This complexity and the size of the third register depend on the precision with which the matrix elements of $A$ are given. A similar procedure can be followed to implement $U_y$.

Using Eq.~\eqref{eq:LCUofA}, we define 
\begin{widetext}
\begin{align}
    \nonumber
    \tilde{H}'(s): = H'(s)\otimes P &= \frac{d+1}{16}X\left(\one-Z\right)\otimes \left( \one - e^{i\pi P}\right) U_x^\dagger U_y S \left( \one - e^{i\pi P}\right) \left( \one + U_{\bar{b}}e^{i\pi P}U_{\bar{b}}^\dagger\right) \\
    \label{eq:LCUofH}
    &+ \frac{d+1}{16}X\left(\one+Z\right)\otimes \left( \one + U_{\bar{b}}e^{i\pi P }U_{\bar{b}}^\dagger\right)\left( \one - e^{i\pi P }\right) U_x^\dagger U_y S \left( \one - e^{i\pi P}\right) \;,
\end{align}
\end{widetext}
where $U_{\bar{b}} = \text{H} \otimes U_b$ is the unitary that prepares the state $\sket{\bar{b}}$ and $\text{H}$ is the Hadamard operation.
This Hamiltonian is a linear combination of 32 unitaries   with equal weights of $(d+1)/16$:
\begin{align}
\label{eq:appHtilde}
    \tilde{H}'(s) =\frac{d+1}{16} \sum_{l=1}^{32} V_l(s) \;.
\end{align}
Note that evolving any state $\ket \psi$ with $H'(s)$
can be done by evolving the state $\ket \psi \otimes \sket{\tilde 0}$
with $\tilde H'(s)$.
Equation~\eqref{eq:appHtilde} is the desired form 
for using the results of Ref.~\cite{BCC+15}.

We want
to simulate the evolution under $\tilde{H}'(s^j)$,
for fixed $s^j$, and evolution time $t^j$. Reference~\cite{BCC+15}
describes a simulation method based on an implementation
of a truncated Taylor series for the evolution operator.
An evolution for time $t^j$ is split into $r$ shorter-time
evolutions (segments) for time $t^j/r$ each. 
Each of these segments is executed by using an algorithm
that implements a linear combination of unitary operations
and by using a technique referred to as oblivious amplitude amplification.

To determine the complexity of the simulation method,
we let $K$ be the truncation order of the Taylor series
approximation of the evolution associated with each segment.
$K$, as well as the number of segments, can be obtained
from the total evolution time under the Hamiltonians $\tilde H'(s^j)$,
which we denote by $T$.
In Ref.~\cite{BCC+15}, it is determined that $K=O(\log(dT/\epsilon)/\log\log(dT/\epsilon))$,
since the sum of the weights of the linear combination 
in Eq.~\eqref{eq:LCUofH} is $O(d)$. The number of segments
for this case is $r=O(dT)$. Since $U_x$ and $U_y$ require $O(1)$ queries
to $A$ each,
it follows that the total number
of $\mathcal{O}_A$ to simulate the evolution for time $T$
is $O(dTK)$. The number of (controlled) $U_b$ and $U^\dagger_b$ 
operations is also $O(dTK)$: each $V_l(s)$ can be applied using, at most, a constant
number of these operations.
The main gate complexity comes from the implementation
of the operations $U_x$ and $U_y$, which is $C_M$, 
and the unitaries $e^{i P}$, which is $O(\log N)=O(n)$.
Thus, the gate complexity of the overall simulation is
$O(dT K (n+C_M))$. Replacing $T$
by the evolution time of our second algorithm, i.e. $T=O(\kappa \log(\kappa)/\epsilon)$,
gives the final query and gate complexities as
$\tilde O(d \kappa/\epsilon)$ and $\tilde O(d \kappa (n+C_M)/\epsilon)$,
respectively. The $\tilde O$ notation hides factors that are 
polylogarithmic in $d\kappa/\epsilon$.

We note that, in the query model,
evolving with $\tilde H'(s)$ is not more complicated 
than evolving with $A$, as the previous Hamiltonian simulation
method yields the same query complexity for both cases.

\end{document}